\documentclass[aps,prl,10pt,twocolumn,superscriptaddress,floatfix,showpacs]
{revtex4-1}

\usepackage{graphicx}
\usepackage{amsfonts}
\usepackage{amssymb}
\usepackage{amsmath}
\usepackage{txfonts}
\usepackage{lipsum}
\usepackage{color}
\usepackage{wasysym}
\usepackage{hyperref}
\usepackage{bbold}

\newcommand{\av}[1]{\ensuremath{\left\langle #1 \right\rangle}}
\newcommand{\qv}{\mathbf{q}}
\newcommand{\kv}{\mathbf{k}}

\usepackage{letltxmacro}
\LetLtxMacro{\oldsqrt}{\sqrt}
\renewcommand{\sqrt}[2][\mkern8mu]{\mkern-6mu\mathop{}\oldsqrt[#1]{#2}}

\begin{document}

\title{
Quantum spin fluctuations and evolution of electronic structure in cuprates
}

\author{E. A. Stepanov}
\email{e.stepanov@science.ru.nl}
\affiliation{\mbox{Radboud University, Institute for Molecules and Materials, 6525AJ Nijmegen, The Netherlands}}
\affiliation{\mbox{Theoretical Physics and Applied Mathematics Department, Ural Federal University, Mira Str. 19, 620002 Ekaterinburg, Russia}}

\author{L. Peters}
\affiliation{\mbox{Radboud University, Institute for Molecules and Materials, 6525AJ Nijmegen, The Netherlands}}

\author{I. S. Krivenko}
\affiliation{\mbox{Department of Physics, University of Michigan, Ann Arbor, Michigan 48109, USA}}

\author{A. I. Lichtenstein}
\affiliation{\mbox{Institute of Theoretical Physics, University of Hamburg, 20355 Hamburg, Germany}}
\affiliation{\mbox{Theoretical Physics and Applied Mathematics Department, Ural Federal University, Mira Str. 19, 620002 Ekaterinburg, Russia}}

\author{M. I. Katsnelson}
\affiliation{\mbox{Radboud University, Institute for Molecules and Materials, 6525AJ Nijmegen, The Netherlands}}
\affiliation{\mbox{Theoretical Physics and Applied Mathematics Department, Ural Federal University, Mira Str. 19, 620002 Ekaterinburg, Russia}}

\author{A. N. Rubtsov}
\affiliation{\mbox{Russian Quantum Center, 143025 Skolkovo,  Russia}}
\affiliation{\mbox{Department of Physics, M.V. Lomonosov Moscow State University, 119991 Moscow, Russia}}

\begin{abstract}
Correlation effects in CuO$_2$ layers give rise to a complicated landscape of collective excitations in high-T$_{\rm c}$ cuprates. Their description requires an accurate account for electronic fluctuations at a very broad energy range and remains a challenge for the theory. Particularly, there is no conventional explanation of the experimentally observed ``resonant'' antiferromagnetic mode, which is often considered to be a mediator of superconductivity. Here we model spin excitations of the hole-doped cuprates in the paramagnetic regime and show that this antiferromagnetic mode is associated with electronic transitions between anti-nodal X and Y points of the quasiparticle band that is pinned to the Fermi level. We observe that upon doping of 7-12\% the electronic spectral weight redistribution leads to the formation of a very stable quasiparticle dispersion due to strong correlation effects. The reconstruction of the Fermi surface results in a flattening of the quasiparticle band at the vicinity of the nodal ${\rm M}\Gamma/2$ point, accompanied by a high density of charge carriers. Collective excitations of electrons between the nodal ${\rm M}\Gamma/2$ and ${\rm XM}/2$ points form the additional magnetic holes state in magnetic spectrum, which protects the antiferromagnetic fluctuation. Further investigation of the evolution of spin fluctuations with the temperature and doping allowed us to observe the incipience of the antiferromagnetic ordering already in the paramagnetic regime above the transition temperature. Additionally, apart from the most intensive low-energy magnetic excitations, the magnetic spectrum reveals less intensive collective spin fluctuations that correspond to electronic processes between peaks of the single-particle spectral function.
\end{abstract}

\maketitle

Despite enormous effort of the theoretical community, electronic structure and quantum spin fluctuations of cuprate compounds remain not well understood~\cite{Anderson1705}. The reason for this lies probably in the fine balance between several competing collective phenomena in these systems, such as superconductivity and the presence of strong charge and spin fluctuations~\cite{RevModPhys.66.763, RevModPhys.84.1383}. The latter is one of the most remarkable properties of cuprates and manifests itself in the antiferromagnetic (AFM) phase at low temperatures in the undoped regime. Moreover, strong electronic correlations imply that collective spin fluctuations are well developed even in the paramagnetic (PM) regime and have a large spin-correlation length. This can be seen as the formation of a Goldstone mode with the frequency proportional to the inverse of the AFM spin-correlation length, and can be observed via the intensity of the spin susceptibility at the $\text{M}=(\pi, \pi)$ point. The correlation length increases with decreasing temperature and the frequency vanishes at the transition temperature forming the AFM ``soft'' mode, as confirmed by the self-consistent spin-wave theory (see Ref.~\cite{PhysRevB.60.1082} and references therein). 

An outstanding property of collective spin excitations in cuprates is 
their extreme robustness against doping. Indeed, in slightly doped cuprate 
compounds the spin-correlation length remains large, and charge carriers move 
in a nearly perfect AFM environment~\cite{RevModPhys.66.763}. The inelastic 
neutron scattering experiments allow to capture  the sharp ``resonance'' in the 
magnon spectrum at the energy of 50-70 meV~\cite{PhysRevLett.70.3490, Dai1344, 
Bourges1234, vignolle2007two, doi:10.1143/JPSJ.81.011007}. This resonant AFM 
mode is present in cuprates within a broad range of temperatures and doping values, and is even proposed as a possible pairing mediator for 
superconductivity~\cite{dahm2009strength, RevModPhys.84.1383}. Various model 
calculations associate this mode either with paramagnetic fluctuations of 
correlated itinerant electrons~\cite{PhysRevB.94.075127, PhysRevB.92.195108} or 
with particle-hole excitations that depend on the band structure of different 
cuprate compounds~\cite{PhysRevX.6.021020, PhysRevB.94.165127}. However, there 
is no conventional understanding of the most distinctive feature of the AFM 
resonance -- why does it remain unchanged in the broad range of doping values?
 
The theoretical description of collective excitations in cuprates requires a 
very advanced approach.  At first glance, the 
Heisenberg~\cite{PhysRevLett.67.3622} and $t$-$J$~\cite{shimahara1992fragility, 
PhysRevB.68.054524} models look suitable for a solution to this problem. However, 
cuprates lie not very deep in the Mott-insulating phase, since the local 
Coulomb interaction $U$ in these systems only slightly exceeds the bandwidth. 
In addition, the presence of the large non-Heisenberg ``ring 
exchange''~\cite{PhysRevB.66.100403} and frustration induced by the 
next-nearest-neighbour hopping $t'$ and nonlocal Coulomb interaction $V$ makes 
a description in terms of localized spins inappropriate. For the same 
reasons, the standard RPA method~\cite{Pines66} is also inapplicable, although 
some attempts in this direction have already been 
made~\cite{PhysRevB.65.014502, PhysRevB.63.054517, PhysRevLett.89.177002}. 
Therefore, the characterization of magnetic fluctuations in terms of 
electronic degrees of freedom requires more elaborated approaches. Some of 
them, such as the quantum Monte-Carlo method~\cite{PhysRevB.92.195108, 
jia2014persistent}, cannot describe collective spin excitations in the most 
interesting physical regime due to the sign problem~\cite{PhysRevB.40.506, 
PhysRevB.41.9301} that appears already far above the transition temperature 
beyond the half-filling. The essential long-range nonlocality of collective 
spin excitations enhanced by the presence of the quasiparticle band at the 
Fermi level of electronic spectrum raises questions about the 
applicability of the extended dynamical mean-field theory 
(EDMFT)~\cite{PhysRevB.52.10295, PhysRevLett.77.3391}. On the other hand, 
the latter is a very efficient description of the Mott-insulating materials 
and can still be used as a basis for further extension of the theory. 
There have been many attempts to go beyond the 
EDMFT~\cite{RevModPhys.90.025003}. However, to our knowledge, the ladder Dual 
Boson (DB) approach~\cite{Rubtsov20121320, PhysRevB.93.045107} is currently the 
only theory that accurately addresses the local and nonlocal collective 
electronic fluctuations in the moderately correlated regime, and remains 
applicable to realistic systems. For example, the DB theory fulfills charge conservation law~\cite{PhysRevLett.113.246407}. Since 
the cuprate compounds show a non-Heisenberg behavior, the magnon-magnon 
interaction plays an extremely important role. Therefore, it should be 
accounted for in the local DB impurity problem via the spin hybridization function 
$\Lambda_{\omega}$, which may violate the spin conservation 
law~\cite{KrienF}.
Recently, it has been shown that the 
latter is still fulfilled if one uses the {\it constant} hybridization function 
$\Lambda$~\cite{KrienF, PhysRevLett.121.037204} in the theory. Therefore, the 
ladder Dual Boson method with the constant hybridization function is a minimal 
approach that correctly accounts for the competing charge and spin excitations 
on an equal footing. 

In this work we consider spin excitations in the two-dimensional $t$-$t'$ extended Hubbard model on a square lattice, which is the simplest model that captures correlation effects in CuO$_2$ layers of cuprates~\cite{ANDERSEN19951573, PhysRevB.53.8751, PhysRevB.92.245113}. Particular parameters of the model are taken to be relevant for the La$_{2}$CuO$_4$ material. Thus, the nearest-neighbor hopping $t=0.3$, the local and nonlocal Coulomb interactions $U=3$ and $V=0.5$, respectively, the direct FM exchange interaction $J^{\rm d}=0.01$ (all units are given in eV) and the next-nearest-neighbor hopping $t'=-0.15t$~\cite{ANDERSEN19951573, PhysRevB.53.8751, PhysRevB.92.245113}. 
It should be noted that there exist several model parametrizations of the cuprate compounds. The mapping of the electronic structure onto the Hubbard model usually leads to a smaller value of the local Coulomb interaction than in the case of the extended Hubbard model. On the other hand, the presence of nonlocal Coulomb interaction in the latter case effectively screens the local Coulomb interaction~\cite{PhysRevLett.111.036601}. Also, the extended Hubbard model considered here enables more accurate description of the nonlocal physics than the Hubbard model.

The model description of cuprate compounds is performed here using the advanced Dual Boson method. The obtained results allow us to explain the phenomenon of robustness of the ``resonant'' mode against doping and to observe a tendency of the system to phase separation between the AFM and conducting holes states. In the undoped case PM spin fluctuations in cuprates show the incipient AFM ``soft'' mode. Finally, apart from the low-energy magnon band, we detect magnetic transitions between peaks (sub-bands) in the single-particle spectral function that are usually observed in resonant inelastic X-ray scattering (RIXS) experiments~\cite{le2011intense, Ghiringhelli1223532, guarise2014anisotropic,PhysRevLett.119.097001, PhysRevB.97.155144} but have not been yet described theoretically.  

\section{Results}

We start the discussion of the obtained results with the most exciting question, namely the existence  of the famous ``resonant'' mode in the spin fluctuation spectrum of cuprates. Since this mode corresponds to a finite frequency, one has to consider collective spin excitations in the paramagnetic regime. Indeed, in the magnetic phase AFM ordering forms the ground state of the system and corresponds to zero frequency. The strongest spin fluctuations in the PM regime emerge in the region close to the phase boundary between the PM and AFM states. Strictly speaking, the long-range order in the two-dimensional systems is allowed only in the ground state, which follows from the Mermin-Wagner theorem. Unfortunately, all modern approaches that provide an approximated solution of the problem based on the momentum space discretization implicitly imply the consideration of a finite system. For the latter case one cannot distinguish between long- and short-range ordering in the system~\cite{RevModPhys.90.025003}. Thus, the transition temperature, which is identified here by the leading eigenvalue $\lambda$ of the Bethe-Salpeter equation for the magnetic susceptibility approaching unity as discussed in~\cite{SM}, corresponds to the disappearance of the short-range order. The latter is referred in the text to as the ``leading magnetic instability''.

\begin{figure}[t!]
\begin{center}
\includegraphics[width=1\linewidth]{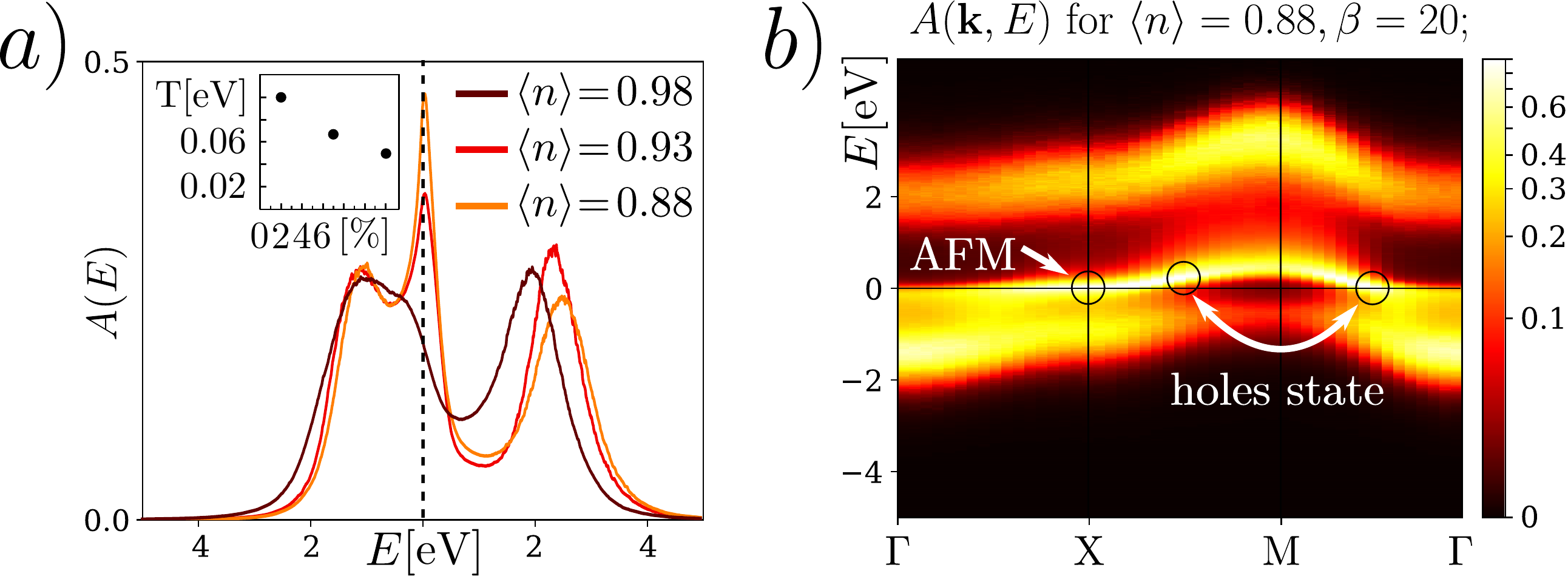}
\end{center}
\vspace{-0.5cm}
\caption{Single particle spectral function $A(E)$ of the extended Hubbard model for cuprates (a) is obtained for the different values of the hole-doping 2, 7 and 12\,\% for $\beta=10$, 15, and 20 eV$^{-1}$, respectively. With the increase of the doping it reveals a sharp peak at the Fermi energy, which corresponds to the existence of the flat band in the momentum space representation of the quasiparticle dispersion $A(\kv,E)$ (b), shown for $\av{n}=0.88$. The inset in (a) shows points in the temperature T[eV] and doping [\%] parameter space where calculations were performed. \label{fig:DOS}}
\end{figure}

Since magnetic fluctuations are by definition collective {\it electronic} excitations the source of the AFM resonant mode should manifest itself already in the {\it single-particle} energy spectrum. 
According to the above discussions, the single particle spectral function $A(E)$ shown in Fig.~\ref{fig:DOS}\,a) is obtained in the normal phase equally close to the phase boundary between the PM and AFM states ($\lambda=0.97\pm0.02$) for different values of the electronic densities $\av{n}=0.98$, 0.93 and 0.88, respectively. The undoped case corresponds to $\av{n}=1$. Note that these results are obtained for different temperatures at which the system is located close to the leading magnetic instability. The corresponding inverse temperatures $\beta$ for these calculations are 10, 15 and 20 eV$^{-1}$, respectively. 

As it is inherent in the Mott insulator, the energy spectrum of the undoped model for cuprates reveals two separated peaks (Hubbard sub-bands) that are located 
below and above the Fermi energy (see Fig.~\ref{fig:DOS}\,a) and~\cite{SM}). 
Upon small doping of $\sim2\,\%$, the two-peak structure of the 
single-particle spectral function transforms to the three-peak structure, where 
the additional quasiparticle resonance appears at the Fermi level splitting off 
from the lower Hubbard band. The further increase of the doping to 7 and 
12\,\% leads to an increase of the quasiparticle peak, which indicates 
the presence of a flat band in the quasiparticle dispersion where excessive 
charge carriers live (see Fig.~\ref{fig:DOS}\,b). Remarkably, after the 
quasiparticle peak appears at the Fermi energy, the flat band at the anti-nodal 
point $\text{X}=(\pi,0)$ is pinned to the Fermi level and does not shift 
anymore with the further increase of the doping. This result is similar to previous theoretical studies of high-T$_{\rm c}$ 
cuprates~\cite{PhysRevLett.89.076401} and Hubbard model on the triangular 
lattice~\cite{PhysRevLett.112.070403}, where the case of the van Hove 
singularity at the Fermi level was considered. Apart from the pinning of the 
Fermi level, we observe that the hole-doping causes the reconstruction of the 
Fermi surface, which manifests itself in the flattening of the energy band at 
the vicinity of the ${\rm M}\Gamma/2=(\pi/2, \pi/2)$ nodal point. 
Redistribution of the spectral weight results in the increased density of holes 
that live around the X and ${\rm M}\Gamma/2$ points as depicted by white arrows 
in Fig.~\ref{fig:DOS}\,b). The rest of the quasiparticle dispersion becomes 
very stable against doping due to strong correlation effects. Thus, the energy 
spectrum is shown here only for one particular case of $\av{n}=0.88$. The other cases of doping are considered in the Supplemental Material~\cite{SM}.

One can also calculate the effective mass renormalization of electrons as $\varepsilon^{*}_{\kv}=Z^{-1}\varepsilon_{\kv}$~\cite{PhysRevB.62.R9283} for different values of doping discussed above.  Here, $\varepsilon_{\kv}$ is the Fourier transform of the hopping matrix
parameterized by $t$ and $t'$. It can be found that in the region close to the magnetic instability the system reveals almost the same renormalization coefficient $Z=4.7\pm0.2$~\cite{SM} for different dopings 2\,\%, 7\,\% and 12\,\%, which additionally confirms the fact that the quasiparticle dispersion becomes stable after it is pinned to the Fermi level. Note that our result for the mass renormalization qualitatively coincides with the experimental value observed in~\cite{dahm2009strength, Putzkee1501657} for another cuprate compound.

\begin{figure}
\begin{center}
\includegraphics[width=1\linewidth]{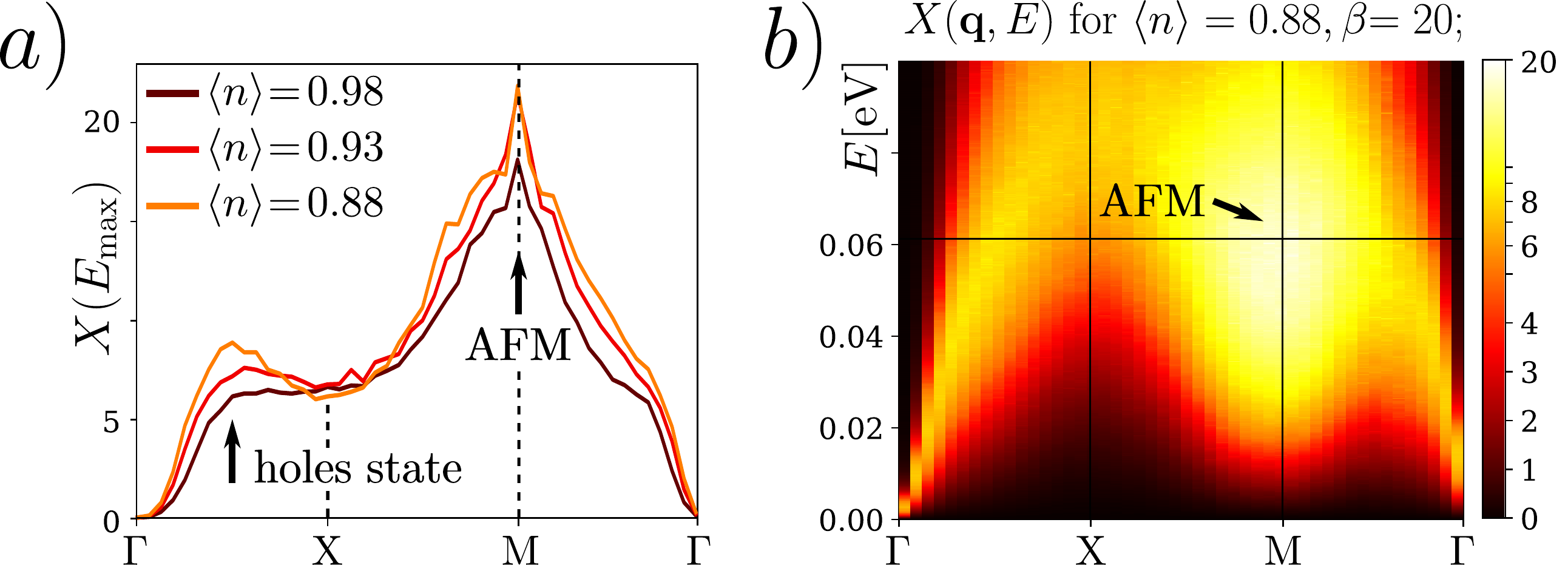}
\end{center}
\vspace{-0.5cm}
\caption{Momentum resolved magnetic susceptibility of the doped extended Hubbard model for cuprates (b) and its cut (a) at the energy $E_{\max}$ that corresponds to a maximum intensity at the M point. The corresponding value of the $E_{\max}$ is almost unchanged and for different hole-doping is $67$ meV (2\,\%), $66$ meV (7\,\%) and $61$ meV (12\,\%). The cut of the magnetic susceptibility reveals two-peaks that correspond to an AFM ordering (M-point) and magnetic holes state ($\Gamma\text{X}/2=(0, \pi/2)$ point). \label{fig:X}}
\end{figure}

Now let us proceed to the two-particle description of the problem and look at 
the low-energy part of the momentum resolved magnetic susceptibility of the 
model shown in Fig.~\ref{fig:X}\,b). Remarkably, the obtained 
dispersion of paramagnons does not change with doping and only reveals 
progressive broadening with an increase of the number of holes in the 
system~\cite{SM} similarly to what has been observed in a recent 
experiment~\cite{PhysRevB.97.155144}. Another distinctive feature of the 
magnetic spectrum that is fortunately captured by the DB method is the high 
intensity at the $\text{M}=(\pi, \pi)$ point. This mode is associated with 
collective AFM fluctuations and is stable against the hole-doping with the 
maximum at the corresponding energies $E_{\rm max}=64\pm3$ meV~\cite{SM}. Since 
specified small differences in the spin fluctuation spectrum are almost 
indistinguishable, the result for the magnetic susceptibility is shown in 
Fig.~\ref{fig:X}\,b) only for one case of $\av{n}=0.88$. Taking into account 
that the presence of doping usually destroys the ordering in the system, the 
result for the magnon dispersion looks counterintuitive at first glance. In 
order to get deeper understanding of this fact, one can look at the cut of the 
magnetic susceptibility at the maximum energy $E_{\rm max}$ shown in 
Fig~\ref{fig:X}\,a) for different values of doping. Then, it becomes 
immediately clear that instead of breaking the AFM ordering, which corresponds 
here to the high peak at the M point, the conducting holes prefer to form their 
own magnetic state that appears as the second peak at the 
$\Gamma\text{X}/2=(\pi/2,0)$ point. Importantly, the height of the minor peak 
grows with the hole-doping, which explains the fact that the AFM mode stays in 
``resonance'' and does not suffer from the existence of the excessive charge 
carriers in the system. A similar momentum-dependent variation of the spectral 
weight of spin fluctuations with doping was also reported 
in~\cite{PhysRevB.97.155144}. The observed picture with no shift of the AFM 
intensity from the M point to an incommensurate position is consistent with the 
scenario of phase separation between the insulating AFM state and conducting 
droplets formed by the excessive charge carriers~\cite{Auslender1982,nagaev1983physics}.

Remarkably, the presence of the observed spin excitations in the doped 
extended Hubbard model for cuprates is reflected in the single-particle spectrum. It is known that in  the undoped regime of the Mott insulator AFM fluctuations are governed by 
Anderson's ``superexchange'' mechanism~\cite{PhysRev.115.2}. Contrarily, 
in the doped case when the quasiparticle band lies at the Fermi energy 
the AFM spin fluctuation arise due to collective excitations of electrons 
between the anti-nodal $\text{X}=(\pi, 0)$ and $\text{Y}=(0, \pi)$ 
points~\cite{PhysRevLett.89.076401, PhysRevLett.112.070403}. 
This fact is also confirmed by the obtained energy 
spectrum (see Fig.~\ref{fig:DOS}\,b)), where the high intensity at the Fermi 
level corresponds to the large density of the charge carriers that live at the 
vicinity of the X point as depicted by the small white arrow. Apart from the 
main AFM fluctuations, the presence of another region of high density of holes, 
appearing at the vicinity of the $\text{M}\Gamma/2=(\pi/2, \pi/2)$ point,
allows an additional magnetic excitation of charge carriers between these two regions as shown by the white curved arrow. This excitation corresponds to the magnetic holes state shown in Fig.~\ref{fig:X} a). Obviously, it is hard to distinguish only two peculiar points of the single-particle spectrum with states above and below the Fermi level that give the main contribution to the specified magnetic excitation, since the spectrum is broadened due to the presence of the large imaginary part of the electronic self-energy. Therefore, there is more than one pair of points that contribute to the magnetic holes state, which is also confirmed by the fact that the corresponding peak in Fig.~\ref{fig:X} a) is relatively wide. However, the momentum space position of the latter allows to estimate the momentum difference between two areas of the single-particle spectrum that are responsible for this magnetic excitation. Looking back at the highest intensity points of the quasiparticle spectrum, one can conclude that the observed minor peak at the $\Gamma\text{X}/2=(\pi/2,0)$ point in the Fig.~\ref{fig:X}\,a) indicates that this excitation happens roughly between the $\text{M}\Gamma/2$ and $\text{XM}/2=(\pi,\pi/2)$ points of the single-particle energy spectrum. Therefore, the redistribution of the quasiparticle weight in addition to the pinning of the quasiparticle spectrum to the Fermi energy allows to keep the single-particle energy spectrum stable against doping, which, in turn, is reflected in the unchanged magnon dispersion.

\begin{figure}[t!]
\begin{center}
\includegraphics[width=1\linewidth]{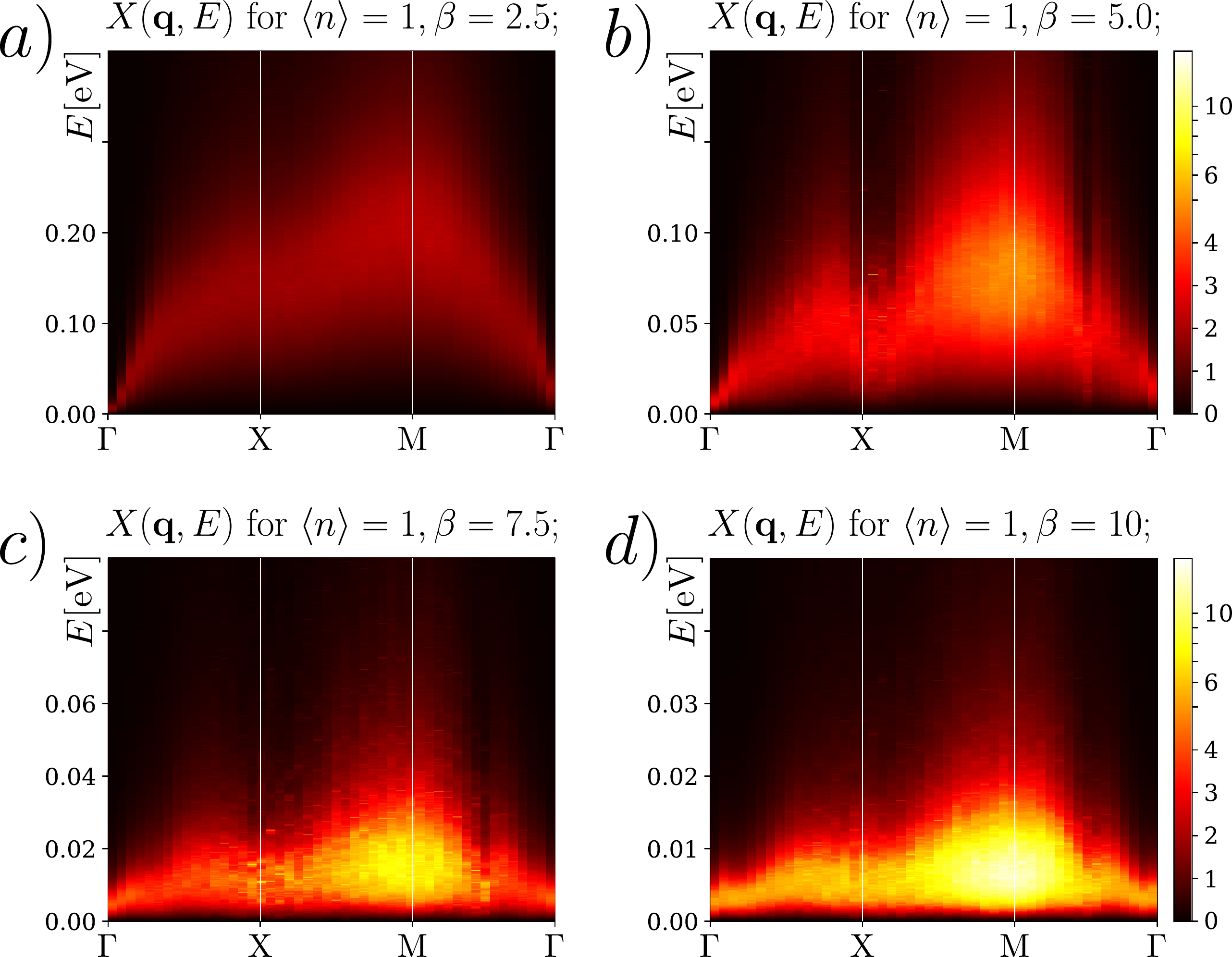}
\end{center}
\vspace{-0.5cm}
\caption{Momentum resolved magnetic susceptibility for the cuprate model for $\beta=2.5$ eV$^{-1}$ (a), 5.0 eV$^{-1}$ (b), 7.5 eV$^{-1}$ (c) and 10 eV$^{-1}$ (d). Intensity at the $\text{M}=(\pi,\pi)$ point corresponds to the formation of the AFM ordering and takes the maximum value at the energy $E_{\max} = 219,\,90,\,18\,\text{and}\,9$ meV, respectively. The latter decreases when approaching the phase transition. \label{fig:undoped}}
\end{figure}

Since our modern approach allows to capture the fingerprint of the AFM ordering already in the paramagnetic phase near the leading magnetic instability, one can go deeper into the PM phase in order to observe the incipience of this fluctuation. Fig.~\ref{fig:undoped} shows the momentum resolved low-energy part of the magnetic susceptibility of the undoped model for different temperatures. The Fig.~\ref{fig:undoped}\,a) corresponds to the case of high temperature ($\beta=2.5$ eV$^{-1}$) and shows a standard paramagnon dispersion.
Lowering the temperature to $\beta=5$ eV$^{-1}$, the characteristic energy scale of spin excitations decreases and the intensity at the M point of the magnon spectrum arises at the energy $E_{\max}=90$ meV (see Fig.~\ref{fig:undoped}\,b)). Since the corresponding energy of the AFM fluctuations is proportional to the inverse of the spin correlation length, it decreases with the temperature as shown in the Fig.~\ref{fig:undoped}\,c) ($\beta=7.5$ eV$^{-1}$) and goes almost to zero approaching the transition temperature at $\beta\simeq10$ eV$^{-1}$ ($\lambda=0.96$) as shown in Fig.~\ref{fig:undoped}\,d). Thus, it can be concluded that the antiferromagnetic mode that forms the ground state of the system in the ordered phase does not appear spontaneously below the transition temperature. On the contrary, it is developed at the finite energy well above the critical temperature already in the paramagnetic phase and ``softens'' approaching the phase boundary, which was also predicted in previous studies (see Ref.~\cite{PhysRevB.60.1082} and references therein).

\begin{figure}[t!]
\begin{center}
\includegraphics[width=1\linewidth]{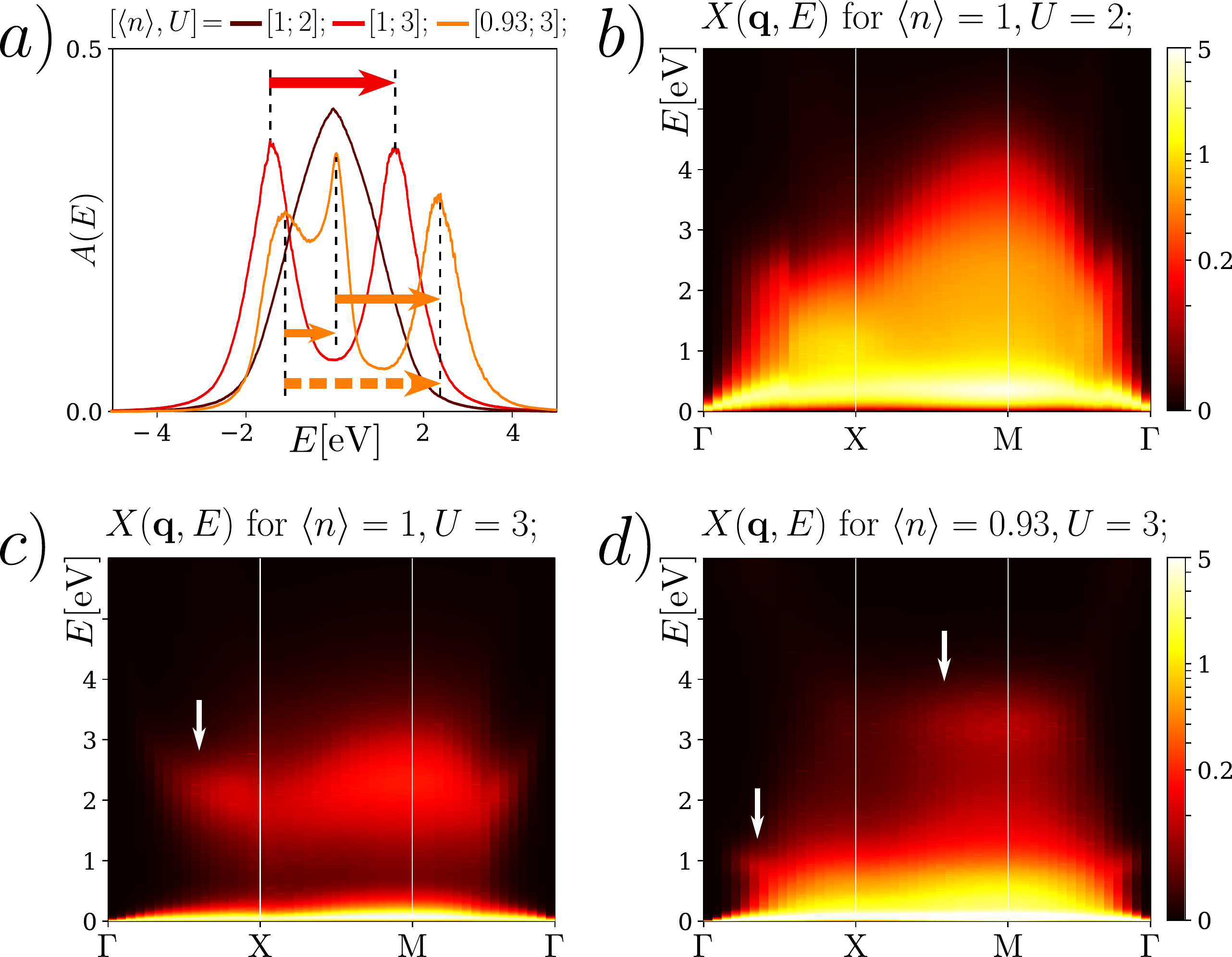}
\end{center}
\vspace{-0.5cm}
\caption{Single particle spectral function (a) and momentum resolved 
magnetic susceptibility in the strongly-correlated metallic $\av{n}=1$, 
$U=2$ eV, $\beta=5$ eV$^{-1}$ (b); Mott-insulating $\av{n}=1$, $U=3$ eV, 
$\beta=5$ eV$^{-1}$ (c); and doped Mott-insulating $\av{n}=0.93$, $U=3$ eV, 
$\beta=15$ eV$^{-1}$ (d) regimes. In addition to the main low-lying mode of 
the high intensity, the magnon spectrum reveals additional one (c) and two (d) 
less pronounced high-energy bands that originate from the magnetic 
excitations between the corresponding peaks in the single particle spectral function 
depicted by the arrows in the top left panel. Energy $E$ is given in the units 
of eV. \label{fig:doped} 
}
\end{figure}

Collective spin excitations of the Mott-insulator that are usually described theoretically are dispersive magnetic excitations that correspond either to the Anderson ``superexchange'' mechanism (in the undoped case), or to the collective electronic processes between the anti-nodal points of the quasiparticle band that lies at the Fermi energy (in the doped case) as discussed above. The characteristic energy of these excitations is of the order of the exchange interaction. In the most general case spin fluctuations are not restricted only to the low-energy magnon band and may reveal additional magnetic excitations. The latter have a completely different energy scale (of the order of the Coulomb interaction in the undoped case) and correspond to the electronic processes between peaks (sub-bands) of the single-particle spectral function. Moreover, they cannot be captured by the most of known theoretical approaches, since they are much less intense than the ``usual'' low-energy ones. 

\begin{figure}[t!]
\begin{center}
\includegraphics[width=0.8\linewidth]{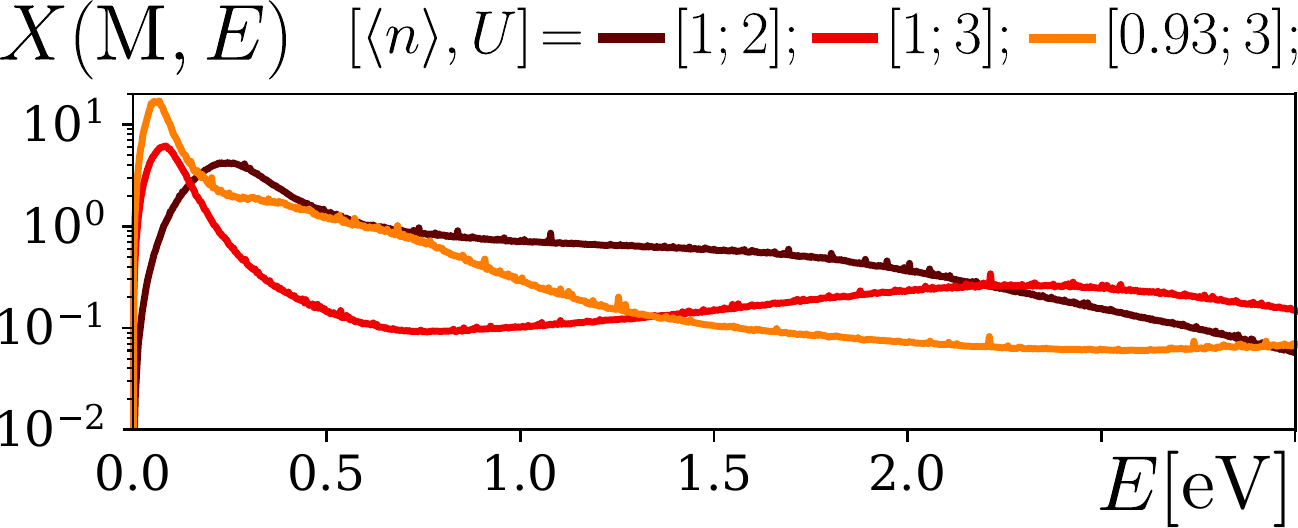}
\end{center}
\vspace{-0.5cm}
\caption{The cut of the momentum resolved 
magnetic susceptibility shown in Figs.~\ref{fig:doped} b)-d) at the M point as the function of the energy. The result is presented in the logarithmic scale. \label{fig:Bandscut} 
}
\end{figure}

In order to study the full spectrum of magnetic fluctuations, let us 
distinguish three cases of interest. First of all, it is worth noting that the considered model for cuprate compounds lies in the region close to the Mott insulator to metal 
phase transition. Reducing the local Coulomb interaction by $1$ eV ($U=2$ eV, 
$\av{n}=1$) gives rise to a single peak in the single particle spectral 
function $A(E)$ in Fig.~\ref{fig:doped}\,a) shifting the material to a metal 
state. In addition, one can specify two more cases ($\av{n}=1$ and 
$\av{n}=0.93$) where the $A(E)$ of the extended Hubbard model for cuprates ($U=3$eV) has a two- 
and three-peak structure, respectively. Corresponding results for the momentum 
resolved magnetic susceptibility shown in Fig.~\ref{fig:doped} reveal one (b), 
two (c) and three (d) magnon bands. The less pronounced high-energy bands in 
Figs.~\ref{fig:doped}\,c) and d) are marked by white arrows. These additional 
bands originate from collective excitations between the specified peaks in the 
single particle spectral function, as depicted by arrows in the 
Fig.~\ref{fig:doped}\,a), similarly to the case of charge 
fluctuations~\cite{PhysRevLett.113.246407}. It is worth mentioning that the process shown in Fig.~\ref{fig:doped}\,a) by the dashed arrow is suppressed, because it occurs between the most distant peaks and does not involve spin 
excitations from the Fermi level, contrary to the other two cases. Therefore, 
the corresponding magnon band is not observed in Fig.~\ref{fig:doped}\,d). For clarity, the cut of the magnetic susceptibility at the M point is shown in Fig.~\ref{fig:Bandscut}. The value of $X(\qv$=M, $E)$ is given in a logarithmic scale in order to distinguish higher-energy bands from the intensive low-energy mode.
Remarkably, the energy scale of these additional magnon bands 
coincides with the RIXS data obtained, for example, in 
Refs.~\cite{Ghiringhelli1223532, guarise2014anisotropic} for another 
cuprate compound. Unfortunately, the RIXS experiment cannot distinguish 
between the charge and spin excitations in the high-energy inter-band 
transitions. Therefore, the corresponding peak shown in these works contains both charge and spin fluctuations, and has the highest amplitude. Thus, the advanced DB scheme allows to capture the higher-energy transitions that are much less intensive than the lower-energy magnon band and to distinguish them from the charge excitations.
To our knowledge, the existence of these high-energy magnetic excitations is reported in the literature for the first time.

\section{Conclusions}

To summarize, in this work electronic properties of the doped extended Hubbard model for  
cuprate compounds in the paramagnetic phase close to the leading magnetic 
instability have been considered. Following the evolution of the electronic 
band structure of cuprates, we have observed that an additional 
quasiparticle band appears at the Fermi level already at the small values of 
doping. Further increase of doping leads to additional flattening of the 
energy band at the vicinity of the nodal $\text{M}\Gamma/2$ point and pinning 
the Fermi level to the anti-nodal points of the quasiparticle band. The 
redistribution of the quasiparticle density results in the spectral weight 
transfer to the vicinity of X (Y) and $\text{M}\Gamma/2$ points, which allows 
the observation of two magnetic modes in the spin fluctuation spectrum. Thus, 
collective electronic excitations between the anti-nodal X and Y points form 
the famous antiferromagnetic ``resonant'' mode, which remains unchanged in a 
broad range of temperatures and dopings. We have shown that protection of 
the AFM resonance is realized simultaneously through the pinning of the 
quasiparticle dispersion to the Fermi energy, and formation of another mode, 
which grows with doping and is located at the $\Gamma\text{X}/2$ point in the 
magnon spectrum. We have discovered that this mode corresponds to collective 
excitations of excessive charge carriers between the nodal $\text{M}\Gamma/2$ 
and anti-nodal XM/2 points. 

The use of the advanced Dual Boson technique allowed us to investigate spin fluctuations in a wide spectral range. Thus, the incipience of the low-energy AFM mode in the undoped model for cuprates is captured in the paramagnetic regime far from the PM to AFM phase transition. This mode softens when approaching the transition temperature and forms the AFM ground state in the broken symmetry phase. The study of higher-energy magnetic fluctuations revealed additional less pronounced magnon bands. We have found that these bands originate from the collective electronic transitions between sub-bands in the quasiparticle energy spectrum and can be captured in the resonant inelastic X-ray scattering experiments~\cite{le2011intense, Ghiringhelli1223532, guarise2014anisotropic,PhysRevLett.119.097001, PhysRevB.97.155144}.  

\section{methods}
The problem of collective excitations in cuprates is addressed here using the Dual Boson theory~\cite{Rubtsov20121320, PhysRevB.93.045107}.  The magnetic susceptibility in the ladder DB approximation is given by the following relation~\cite{PhysRevLett.121.037204} 
\begin{align}
\left[X^{\rm ladd}_{\qv\omega}\right]^{-1} = J^{\rm d}_{\qv} + \Lambda + \left[X^{\rm DMFT}_{\qv\omega}\right]^{-1},
\end{align}
where $X^{\rm DMFT}_{\qv\omega}$ is the DMFT-like~\cite{PhysRevLett.62.324, 
RevModPhys.68.13} magnetic susceptibility written in terms of the local 
two-particle irreducible four-point vertices and lattice Green's functions. The 
latter is dressed only in the local self-energy and is given by the usual EDMFT 
relation~\cite{PhysRevB.52.10295, PhysRevLett.77.3391}. The single- and 
two-particle spectral functions are obtained, respectively, from the lattice 
Green's function and magnetic susceptibility by a stochastic optimization 
method for analytic continuation~\cite{SOM,PhysRevB.62.6317}. 
The details of calculations can be found in~\cite{SM}.

The effective mass renormalization $Z$ of electrons can be found as $\varepsilon^{\ast}_{\kv}=Z^{-1}\varepsilon_{\kv}$, where the coefficient $Z$ reads~\cite{PhysRevB.62.R9283}
\begin{align}
Z = 1-\left. \frac{d \Sigma_{E}}{d E} \right|_{E=0},
\end{align}
since in the ladder DB approximation the electronic self-energy $\Sigma_{E}$ does not depend on momentum $\kv$. Importantly, the calculation of the renormalization coefficient does not require the analytical continuation procedure. The result for the electronic self-energy can be found in~\cite{SM}.

The data that support the findings of this study are available from the corresponding author upon reasonable request.

Authors thank Nigel Hussey for inspiring discussions. We also thank Hartmut Hafermann for providing the impurity solver~\cite{HAFERMANN20131280} based on the ALPS libraries~\cite{1742-5468-2011-05-P05001}, and Erik van Loon, Friedrich Krien and Arthur Huber for the help with the Dual Boson implementation.  

The Authors declare no Competing Financial or Non-Financial Interests.

All authors discussed the results and contributed to the preparation of the manuscript. 

E.A.S. and M.I.K. would like to thank the support of NWO via Spinoza Prize and of ERC Advanced Grant 338957 FEMTO/NANO. Also, E.A.S. and M.I.K. acknowledge the Stichting voor Fundamenteel Onderzoek der Materie (FOM), which is financially supported by the Nederlandse Organisatie voor Wetenschappelijk Onderzoek (NWO). I.S.K. acknowledges support from U.S. Department of Energy, Office of Science via Grant No. DOE ER 46932. A.I.L. acknowledges support from the excellence  cluster ``The Hamburg Centre for Ultrafast Imaging - Structure, Dynamics and Control of Matter at the Atomic Scale'' and North-German Supercomputing Alliance (HLRN) under the Project No. hhp00040. The contribution of A.I.L. and A.N.R. was funded by the joint Russian Science Foundation (RSF)/DFG Grant No. 16-42-01057 / LI 1413/9-1.

\bibliography{Ref}

\clearpage
\onecolumngrid

\begin{center}
{\Large Supplemental Material for \\
Quantum spin fluctuations and evolution of electronic structure in cuprates}
\end{center}

\section{Action}

The action of the considered extended Hubbard model written in momentum space has the following form
\begin{align}
\label{eq:Hamiltonian}
{\cal S} = -\sum_{\kv, \nu, \sigma}c^{*}_{\kv\nu\sigma}\left[i\nu+\mu-\varepsilon^{\phantom{*}}_{\kv}\right]
\,c^{\phantom{\dagger}}_{\kv\nu\sigma}
+ U\sum\limits_{\qv, \omega}n^{*}_{\qv\omega\uparrow}n^{\phantom{*}}_{\qv\omega\downarrow} 
+\frac12\sum\limits_{\qv, \omega, \varsigma}V^{\,\varsigma}_{\qv}\rho^{*\,\varsigma}_{\qv\omega}\,\rho^{\,\varsigma}_{\qv\omega}. 
\end{align}
Here, $c^{*}_{\kv\nu\sigma}$ ($c_{\kv\nu\sigma}$) are Grassmann variables corresponding to creation (annihilation) of an electron with momentum $\kv$, fermionic Matsubara frequency $\nu$ and spin $\sigma$. $\varepsilon_{\kv}$ is the Fourier transform of the nearest-neighbor (NN) $t$ and next-NN $t'$ hopping amplitudes. The label $\varsigma=\{c,s\}$ depicts charge $c$ and spin $s=\{x,y,z\}$ degrees of freedom, so that $U$ and $V^{\,c}_{\qv}=V_{\qv}$ describe local and nonlocal parts of the Coulomb interaction respectively, and $V^{\,s}_{\qv}=-J^{\rm d}_{\qv}/2$ is the nonlocal direct ferromagnetic exchange interaction. 
Here, we also introduce bosonic variables $\rho^{\,\varsigma}_{\qv\omega}=~n^{\,\varsigma}_{\qv\omega}-~\av{n^{\,\varsigma}_{\qv\omega}}$, where $n^{\,\varsigma}_{\qv\omega}=\sum_{\kv\nu\sigma\sigma'}c^{*}_{\kv+\qv,\nu+\omega,\sigma}\sigma^{\,\varsigma}_{\sigma\sigma'}c^{\phantom{*}}_{\kv,\nu,\sigma'}$ is the charge (spin) density with momentum $\qv$ and bosonic Matsubara frequency $\omega$ and $\sigma^{\,c\,(s)}=I\,(\sigma)$ is the unit (Pauli) matrix for the charge (spin) channels, respectively.

The description of collective excitations is given here within the ladder Dual Boson theory~~\cite{Rubtsov20121320, PhysRevB.90.235135, PhysRevB.93.045107, PhysRevB.94.205110}, which 
implies exact solution of the corresponding local impurity problem
\begin{align}
\label{eq:imp}
{\cal S}_{\rm imp} = &-\sum\limits_{\nu, \sigma}c^{*}_{\nu\sigma}
\left[i\nu+\mu-\Delta^{\phantom{*}}_{\nu}\right]c^{\phantom{*}}_{\nu\sigma}
+ U\sum\limits_{\omega}n^{*}_{\omega\uparrow}n^{\phantom{*}}_{\omega\downarrow} 
+\frac12\sum\limits_{\omega, \varsigma}\Lambda^{\,\varsigma}_{\omega}\,\rho^{*\,\varsigma}_{\omega}\,\rho^{\,\varsigma}_{\omega},
\end{align}
where the fermionic $\Delta_{\nu}$ and bosonic $\Lambda^{\varsigma}_{\omega}$ hybridization functions are introduced similarly to the EDMFT~\cite{PhysRevB.52.10295, PhysRevLett.77.3391, PhysRevB.61.5184, PhysRevLett.84.3678, PhysRevB.63.115110, PhysRevLett.90.086402} in order to effectively account for nonlocal excitations and have to be determined self-consistently. Note that the same functions have to be excluded from the remaining nonlocal part of the action 
\begin{align}
{\cal S}_{\rm rem} = 
-\sum\limits_{\kv, \nu, \sigma}c^{*}_{\kv\nu\sigma}
\left[\Delta^{\phantom{*}}_{\nu}-\varepsilon^{\phantom{*}}_{\kv}\right]
c^{\phantom{*}}_{\bf{k}\nu\sigma}
+\frac12\sum\limits_{\qv, \omega, \varsigma} \left(V^{\,\varsigma}_{\qv} - \Lambda^{\,\varsigma}_{\omega}\right) \rho^{*\,\varsigma}_{\qv\omega}\,\rho^{\,\varsigma}_{\qv\omega},
\end{align}
so that the total lattice problem ${\cal S}=\sum_{i}{\cal S}^{(i)}_{\rm 
imp}+{\cal S}_{\rm rem}$ is unchanged. Inclusion of bosonic hybridization 
functions is important and leads to a great improvement of results already at 
the dynamical mean-field level~\cite{PhysRevB.90.235135, PhysRevB.94.205110}. 
Nevertheless, this procedure has some hidden difficulties. As it was shown 
recently, while the bosonic hybridization function $\Lambda^{c}_{\omega}$ in 
the charge channel performs rather well, the account for the same kind of 
frequency dependent function $\Lambda^{s}_{\omega}$ in the spin channel leads 
to the changed Ward identity and thus breaks the local spin conservation at the 
impurity level~\cite{PhysRevB.96.075155}. Also, the inclusion of the bosonic 
hybridization function in the spin channel drastically complicates solution of 
the impurity problem and is often associated with the sign problem. Whereas the 
solution of this issue in the single-band case was recently 
proposed~\cite{PhysRevB.87.125102}, an application of this method to realistic 
multiorbital systems is extremely complicated and is not done yet. However, 
there is still one particular form of the bosonic hybridization function, which 
does not violate local conservation laws, stays almost undiscussed. Indeed, 
when the bosonic hybridization for the spin channel is approximated by a 
constant function in the frequency space $\Lambda^{s}_{\omega}\to\Lambda^{s}$, 
the local Ward identity remains unchanged and conservation laws are 
fulfilled~\footnote{This conclusion can be made when replacing the frequency 
dependent hybridization function in the spin channel by a constant in 
derivation of Ward identity in~\cite{PhysRevB.96.075155}}. 

\section{Variation of the impurity problem}

The important consequence of introducing of retarded interactions is that every variation $\delta\Delta_{\nu}$ and $\delta\Lambda_{\omega}^{\,\varsigma}$ doesn't change the total action and, as a consequence, the partition function. It is also possible to vary retarded interactions in such a way that the impurity problem~\eqref{eq:imp} remains unchanged as well. According to above discussions, let us assume that $\delta\Delta$ and $\delta\Lambda^{\,\varsigma}$ are {\it constant} variations of fermionic and bosonic hybridization functions. Therefore, the variation of the impurity action reads
\begin{align}
2\,\delta{\cal S}_{\rm imp} &= 2\,\delta\Delta\sum_{\nu\sigma}c^{*}_{\nu\sigma}c^{\phantom{*}}_{\nu\sigma} + \delta\Lambda^{c}\sum_{\omega}\rho^{*\,c}_{\omega}\,\rho^{\,c}_{\omega} + \delta\Lambda^{s}\sum\limits_{s,\omega}\rho^{*\,s}_{\omega}\,\rho^{\,s}_{\omega}.\\
&=
2\,\delta\Delta\sum_{\nu\sigma}c^{*}_{\nu\sigma}c^{\phantom{*}}_{\nu\sigma} + 2\,\delta\Lambda^{c}\sum_{\omega}n^{*}_{\omega\uparrow}n^{\phantom{*}}_{\omega\downarrow} - 
\delta\Lambda^{c} \sum_{\nu\sigma}c^{*}_{\nu\sigma}c^{\phantom{*}}_{\nu\sigma} + \delta\Lambda^{c}\av{n} - 
6\,\delta\Lambda^{s}\sum\limits_{\omega}n^{*}_{\omega\uparrow} n^{\phantom{*}}_{\omega\downarrow} + \delta\Lambda^{s} \sum_{\nu\sigma}c^{*}_{\nu\sigma}c^{\phantom{*}}_{\nu\sigma} \notag\\
&=
\Big(2\,\delta\Delta-\delta\Lambda^{c}+\delta\Lambda^{s}\Big) \sum_{\nu\sigma}c^{*}_{\nu\sigma}c^{\phantom{*}}_{\nu\sigma} +
2\,\Big(\delta\Lambda^{c} - 3\delta\Lambda^{s}\Big) \sum_{\omega}n^{*}_{\omega\uparrow}n^{\phantom{*}}_{\omega\downarrow} + \delta\Lambda^{c}\av{n}, \notag
\end{align}
where we considered an anisotropic case of spin fluctuations 
$\delta\Lambda^{x}=\delta\Lambda^{y}=\delta\Lambda^{z}=\delta\Lambda^{s}$. 
Since the impurity problem is assumed to be unchanged under these 
transformations, one gets the following relations for the introduced 
variations
\begin{align}
&\delta\Lambda^{c} = 3\,\delta\Lambda^{s}, \label{eq:Var1}\\
&\delta\Delta = \frac{\delta\Lambda^{c}-\delta\Lambda^{s}}{2}=\delta\Lambda^{s}.
\label{eq:Var2}
\end{align}
Here, the last relation describes a constant shift of the chemical potential $\mu$. Thus, the total variation of the impurity action is
\begin{align}
\delta{\cal S}_{\rm imp} = \delta\Lambda^{c}\av{n}/2,
\end{align}
which is just a constant shift of the energy that does not affect the calculation of local observables. Then, based on the above transformation, the initial action~\eqref{eq:imp} of the impurity model can be simplified as
\begin{align}
\label{eq:imp2}
{\cal S}_{\rm imp} 
&= -\sum\limits_{\nu, \sigma}c^{*}_{\nu\sigma}
\left[i\nu+\mu-\Delta^{\phantom{*}}_{\nu}\right]c^{\phantom{*}}_{\nu\sigma}
+ U\sum\limits_{\omega}n^{*}_{\omega\uparrow}n^{\phantom{*}}_{\omega\downarrow} 
+\frac12\sum\limits_{\omega, \varsigma}\Lambda^{\,\varsigma}_{\omega}\,\rho^{*\,\varsigma}_{\omega}\,\rho^{\,\varsigma}_{\omega},\\
&= -\sum\limits_{\nu,\sigma}c^{*}_{\nu\sigma}\left[i\nu+\mu-\Delta_{\nu}\right]c^{\phantom{*}}_{\nu\sigma} + U\sum_{\omega}n^{*}_{\omega\uparrow}n^{\phantom{*}}_{\omega\downarrow} 
+\frac12\sum\limits_{\omega}\left[\Lambda^{c}_{\omega} - 3\Lambda^{s}\right]\rho^{*\,c}_{\omega}\rho^{\,c}_{\omega} - \Lambda^{s}\sum_{\nu\sigma}c^{*}_{\nu\sigma}c^{\phantom{*}}_{\nu\sigma} \notag\\
&= -\sum\limits_{\nu,\sigma}c^{*}_{\nu\sigma}\left[i\nu+\tilde{\mu}-\Delta_{\nu}\right]c^{\phantom{*}}_{\nu\sigma} 
+U\sum_{\omega}n^{*}_{\omega\uparrow}n^{\phantom{*}}_{\omega\downarrow} + \frac12\sum\limits_{\omega} \Lambda^{0}_{\omega}\, \rho^{*\,c}_{\omega}\,\rho^{\,c}_{\omega}, \notag
\end{align}
where $\delta\Lambda^{s}=-\Lambda^{s}$, $\tilde{\mu}=\mu+\Lambda^{s}$ and $\Lambda^{0}=\Lambda^{c}_{\omega}-3\Lambda^{s}$. 

If the charge hybridization function is also taken as a constant $\Lambda^{c}_{\omega}=\Lambda^{c}$, then the action~\eqref{eq:imp2} simplifies to 
\begin{align}
\label{eq:imp3}
{\cal S}_{\rm imp} 
=-\sum\limits_{\nu,\sigma}c^{*}_{\nu\sigma}\left[i\nu+\mu'-\Delta_{\nu}\right]c^{\phantom{*}}_{\nu\sigma} 
+U'\sum_{\omega} n^{*}_{\omega\uparrow}n^{\phantom{8}}_{\omega\downarrow},
\end{align}
where $\mu'=\mu+(\Lambda^{c}-\Lambda^{s})/2$ and $U' = U + \Lambda^{c}-3\Lambda^{s}$.

Therefore, all local observables of impurity model~\eqref{eq:imp} can be 
calculated using simpler local problems written in the 
EDMFT~\eqref{eq:imp2} and DMFT~\eqref{eq:imp3} forms. It turns out that this 
approximation is very attractive for numerical calculations, since the 
inclusion of the spin channel in the impurity problem~\eqref{eq:imp} does not 
require additional implementation, since the simplified actions~\eqref{eq:imp2} 
and~\eqref{eq:imp3} contain only the charge degrees of freedom.

\section{Variation of the lattice problem}
The partition function of the initial problem is given by the following relation
\begin{align}
{\cal Z}=\int D[c] \, e^{-{\cal S}}.
\end{align}
According to the usual formulation of the Dual Boson theory~\cite{Rubtsov20121320, PhysRevB.90.235135, PhysRevB.93.045107, PhysRevB.94.205110}, one can perform Hubbard--Stratonovich transformations of the nonlocal part of the action ${\cal S}_{\rm rem}$ and introduce new ${\it dual}$ variables $f^{*},f,\phi^{\,\varsigma}$ as
\begin{align}
\exp\left\{\,\sum\limits_{{\bf k}\nu\sigma} c^{*}_{{\bf k}\nu\sigma}[\Delta_{\nu\sigma}-\varepsilon_{\bf k}]c^{\phantom{*}}_{{\bf k}\nu\sigma}\right\} &= {\cal D}_{f}
\int D[f]\,\exp\left\{-\sum\limits_{{\bf k}\nu\sigma}\left( f^{*}_{{\bf k}\nu\sigma}[\Delta_{\nu\sigma}-\varepsilon_{\bf k}]^{-1}f^{\phantom{*}}_{{\bf k}\nu\sigma} + c^{*}_{\nu\sigma}f^{\phantom{*}}_{\nu\sigma} + f^{*}_{\nu\sigma}c^{\phantom{*}}_{\nu\sigma}\right)\right\},\notag\\
\exp\left\{\,\frac12\sum\limits_{\qv\omega} \rho^{*\,\varsigma}_{{\bf q}\omega}[\Lambda^{\,\varsigma}_{\omega}-V^{\,\varsigma}_{\qv}]\rho^{\,\varsigma}_{\qv\omega}\right\} &= {\cal D}_{\varsigma}
\int D[\phi^{\,\varsigma}]\,\exp\left\{-\frac12\sum\limits_{\qv\omega}\left( \phi^{*\,\varsigma}_{\qv\omega}
[\Lambda^{\,\varsigma}_{\omega}-V^{\,\varsigma}_{\qv}]^{-1}\phi^{\,\varsigma}_{\qv\omega} + \rho^{*\,\varsigma}_{\omega}
\phi^{\,\varsigma}_{\omega} + \phi^{*\,\varsigma}_{\omega}\rho^{\,\varsigma}_{\omega}\right)\right\},\notag
\end{align}
where ${\cal D}_{f} = {\rm det}[\Delta_{\nu\sigma}-\varepsilon_{\bf k}]$ and ${\cal D}^{-1}_{\,\varsigma} = \sqrt{{\rm det}[\Lambda^{\varsigma}_{\omega}-V^{\,\varsigma}_{\bf q}]}$.
Rescaling the fermionic field as $f_{{\bf k}\nu\sigma}\to{}f_{{\bf 
k}\nu\sigma}g^{-1}_{\nu\sigma}$ and bosonic field as 
$\phi^{\,\varsigma}_{\qv\omega}\to{}\phi^{\,\varsigma}_{\qv\omega}\alpha^{\,\varsigma~-1}_{\omega}$, we transform the initial problem~\eqref{eq:imp} to ${\cal S} = \sum_{i}{\cal S}^{(i)}_{\rm imp} + {\cal S}_{\rm DB}$, where
\begin{align}
{\cal S}_{\rm DB} = 
&-\sum_{\kv,\nu,\sigma} f^{*}_{\kv\nu\sigma}g^{-1}_{\nu\sigma} (\varepsilon_{\kv}-\Delta_{\nu})^{-1} g^{-1}_{\nu\sigma}
f^{\phantom{*}}_{\kv\nu\sigma} 
+\sum_{\kv,\nu,\sigma} \left[ c^{*}_{\kv\nu\sigma}g^{-1}_{\nu\sigma}f^{\phantom{*}}_{\kv\nu\sigma} + 
f^{*}_{\kv\nu\sigma}g^{-1}_{\nu\sigma}c^{\phantom{*}}_{\kv\nu\sigma} \right] \notag\\
&-\frac12\sum_{\qv,\omega,\varsigma} \phi^{*\,\varsigma}_{\qv\omega}\alpha^{\,\varsigma~-1}_{\omega} (V^{\,\varsigma}_{\qv}-\Lambda^{\varsigma}_{\omega})^{-1}\alpha^{\,\varsigma~-1}_{\omega} \phi^{\,\varsigma}_{\qv\omega}
+\frac12\sum_{\qv,\omega} \left[ \rho^{*\,\varsigma}_{\qv\omega}\alpha^{\,\varsigma~-1}_{\omega}\phi^{\,\varsigma}_{\qv\omega} + 
\phi^{*\,\varsigma}_{\qv\omega}\alpha^{\,\varsigma~-1}_{\omega}\rho^{\,\varsigma}_{\qv\omega} \right]. \notag
\end{align}

Let us now consider same variations of the lattice problem
\begin{align}
\frac{\delta{\cal Z}}{\cal Z} &= 
\frac{1}{\cal Z}\,\delta\left[{\cal D}_{f}{\cal D}_{c}{\cal D}_{s}\int D[c, f, \phi^{c}, \phi^{s}] \,
e^{-{\cal S}_{\rm DB}}e^{- \sum_{i}{\cal S}^{(i)}_{\rm imp}}  \right] \\
&= 
\frac{\delta{}{\cal D}_{f}}{{\cal D}_{f}} + \frac{\delta{}{\cal D}_{c}}{{\cal D}_{c}} + \frac{\delta{}{\cal D}_{s}}{{\cal D}_{s}} + 
\frac{{\cal D}_{f}{\cal D}_{c}{\cal D}_{s}}{\cal Z}\int D[c, f, \phi^{c}, \phi^{s}] 
\left[-\delta{}{\cal S}_{\rm DB}\right]e^{-{\cal S}} + 
\frac{{\cal D}_{f}{\cal D}_{c}{\cal D}_{s}}{\cal Z}\int D[c, f, \phi^{c}, \phi^{s}] 
\left[-\sum\limits_{i}\delta{}{\cal S}^{i}_{\rm imp}\right]e^{-{\cal S}} \notag
\end{align}
that do not change the partition function ${\cal Z}$ as discussed above. Since the previously obtained total variation of the impurity action is $\delta{\cal S}_{\rm imp}=\delta\Lambda^{c}\av{n}/2$, one gets the following expression
\begin{align}
0 = \delta\Lambda^{c}\av{n} &+ 
2\delta\Delta\sum_{\kv,\nu,\sigma}
\left\{\left[\Delta_{\nu\sigma}-\varepsilon_{\kv}\right]^{-1} g^{-1}_{\nu\sigma}\av{f^{*}_{\kv\nu\sigma}
f^{\phantom{*}}_{\kv\nu\sigma}}_{\rm latt}g^{-1}_{\nu\sigma} \left[\Delta_{\nu\sigma}-\varepsilon_{\kv}\right]^{-1} -
\left[\Delta_{\nu\sigma}-\varepsilon_{\kv}\right]^{-1}\right\}\notag\\ 
&+\delta\Lambda^{c}\sum_{\qv,\omega}
\left\{\left[\Lambda^{c}_{\omega}-V_{\qv}^{c}\right]^{-1} \alpha^{-1\,c}_{\omega}\av{\phi^{*\,c}_{\qv\omega}
\phi^{c}_{\qv\omega}}_{\rm latt}\alpha^{-1\,c}_{\omega} \left[\Lambda^{c}_{\omega}-V_{\qv}^{c}\right]^{-1} -
\left[\Lambda^{c}_{\omega}-V_{\qv}^{c}\right]^{-1}\right\}\notag\\
&+\delta\Lambda^{s}\sum_{\qv,\omega,s}
\left\{\left[\Lambda^{s}_{\omega}-V_{\qv}^{s}\right]^{-1} \alpha^{-1\,s}_{\omega}\av{\phi^{*\,s}_{\qv\omega}
\phi^{s}_{\qv\omega}}_{\rm latt}\alpha^{-1\,s}_{\omega} \left[\Lambda^{s}_{\omega}-V_{\qv}^{s}\right]^{-1} -
\left[\Lambda^{s}_{\omega}-V_{\qv}^{s}\right]^{-1}\right\}.
\end{align}
Using the exact relation between the lattice and dual quantities~\cite{Rubtsov20121320, PhysRevB.90.235135, PhysRevB.93.045107, PhysRevB.94.205110} and connections between variations of hybridization functions~\eqref{eq:Var1} and~\eqref{eq:Var2}, one gets the following analytical expression for the ``Pauli principle''
\begin{align}
3\sum_{\qv,\omega}X^{c}_{\qv\omega} + \sum_{\qv,\omega,s}X^{s}_{\qv\omega}&=3\av{n}-2\sum_{\kv,\nu,\sigma}G_{\kv\nu\sigma} = \av{n}.
\end{align}
Here $X_{\qv\omega}$ and $\chi_{\omega}$ are the lattice and impurity susceptibilities, respectively. Therefore, if one takes the 
\begin{align}
\label{eq:SC}
\sum_{\qv,\omega}X^{\varsigma}_{\qv\omega} = \sum_{\omega}\chi^{\varsigma}_{\omega}
\end{align} 
self-consistency condition on the hybridization function $\Lambda^{\varsigma}$, 
the ``Pauli principle'' is fulfilled automatically, because the impurity 
problem is solved numerically exactly and the following relation
\begin{align}
3\sum_{\omega}\chi^{c}_{\omega} + \sum_{s,\omega}\chi^{s}_{\omega} &= \av{n}
\end{align}
is correct by definition. 

\section{Calculation of observables}

\begin{figure}[!t]
\begin{center}
\includegraphics[width=1\linewidth]{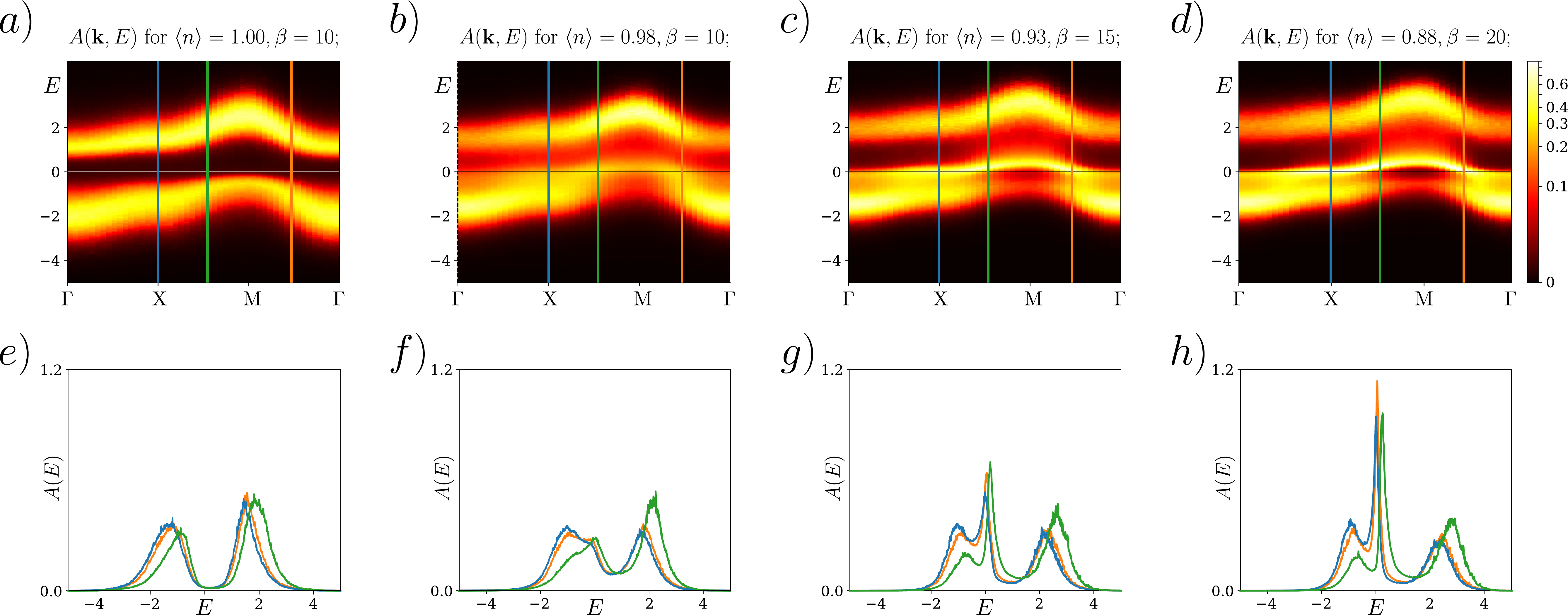}
\end{center}
\vspace{-0.3cm}
\caption{Evolution of the energy spectrum $A(\qv,E)$ of the La$_2$CuO$_4$ (top row) and its cut at X, XM/2 and $\text{M}\Gamma/2$ (bottom row) obtained in the paramagnetic regime close to the PM to AFM phase boundary for different values of hole-doping (from left to right) $\av{n}=1$, 0.98, 0.93, and 0.88, respectively. The energy $E$ is given in the units of eV.  \label{fig:SM_Green}}
\end{figure}

\begin{figure}[b!]
\begin{center}
\includegraphics[width=0.75\linewidth]{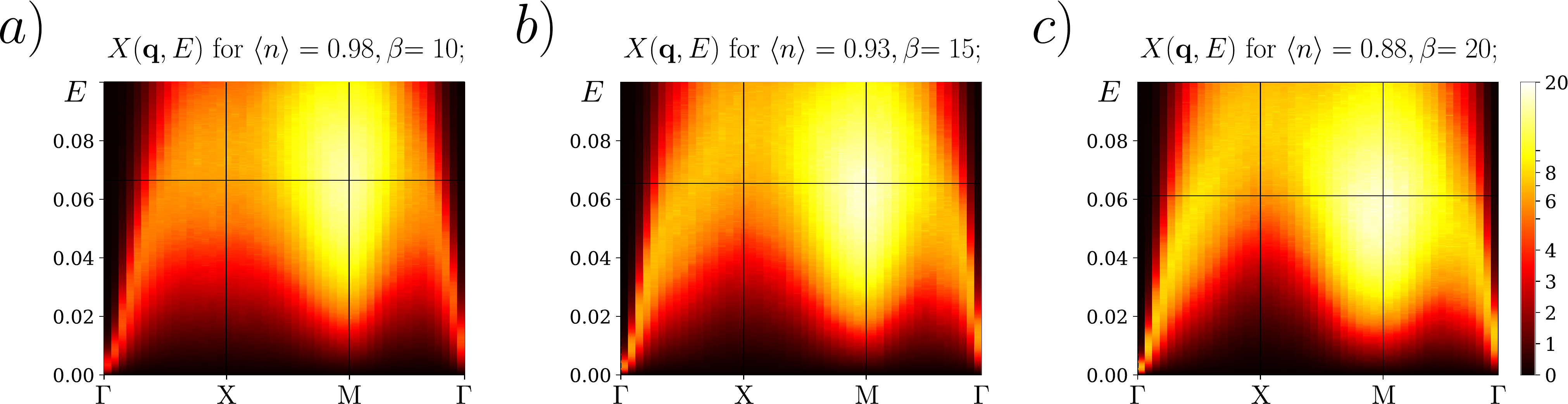}
\end{center}
\vspace{-0.3cm}
\caption{Momentum resolved magnetic susceptibility of the cuprate model obtained in the paramagnetic regime close to the PM to AFM phase boundary for different values of hole-doping (from left to right) $\av{n}=0.98$, $\av{n}=0.93$ and $\av{n}=0.88$. The energy $E$ is given in the eV. \label{fig:SM_X}}
\end{figure}

All lattice quantities of the considered problem can be obtained following the standard Dual Boson scheme~\cite{Rubtsov20121320, PhysRevB.90.235135, PhysRevB.93.045107, PhysRevB.94.205110} with the only one difference in the self-consistency condition~\eqref{eq:SC} on the constant bosonic hybridization function $\Lambda^{\varsigma}$. In our work we restrict ourselves to the ladder Dual Boson description of collective excitations. Therefore, the lattice self-energy $\Sigma_{\kv\nu\sigma}$ is approximated by that of the local impurity problem~\eqref{eq:imp} and the nonlocal contribution is omitted in order to obey charge and spin conservation laws. Then, the lattice Green's function is equal to the EDMFT Green's function $G_{\kv\nu\sigma}$ and the magnetic susceptibility can be written in the following form~\cite{PhysRevLett.121.037204}
\begin{align}
\left[X^{\rm ladd}_{\qv\omega}\right]^{-1} = J^{\rm d}_{\qv} + \Lambda + \left[X^{\rm DMFT}_{\qv\omega}\right]^{-1},
\label{eq:X_DB}
\end{align}
where $X^{\rm DMFT}_{\qv\omega}$ is the DMFT-like~\cite{PhysRevLett.62.324, 
RevModPhys.68.13} magnetic susceptibility written in terms of the local 
two-particle irreducible four-point vertices and lattice Green's functions. 
Numerical calculations of the Green's function and susceptibility are performed 
on the $32\times32$ lattice. Number of $\bf{k}$ points in the Brillouin Zone is 
the same as for the lattice sites, namely $32\times32$. Number of fermionic 
Matsubara frequencies is 36, which is twice larger than the bosonic one. The 
single- and two-particle spectral functions are shown in 
Fig.~\ref{fig:SM_Green} and Fig.~\ref{fig:SM_X}, respectively, and obtained 
from the lattice Green's function and magnetic susceptibility by the 
stochastic optimization method for analytic continuation~\cite{SOM, 
PhysRevB.62.6317}. 

\begin{figure}[h!]
\begin{center}
\includegraphics[width=0.3\linewidth]{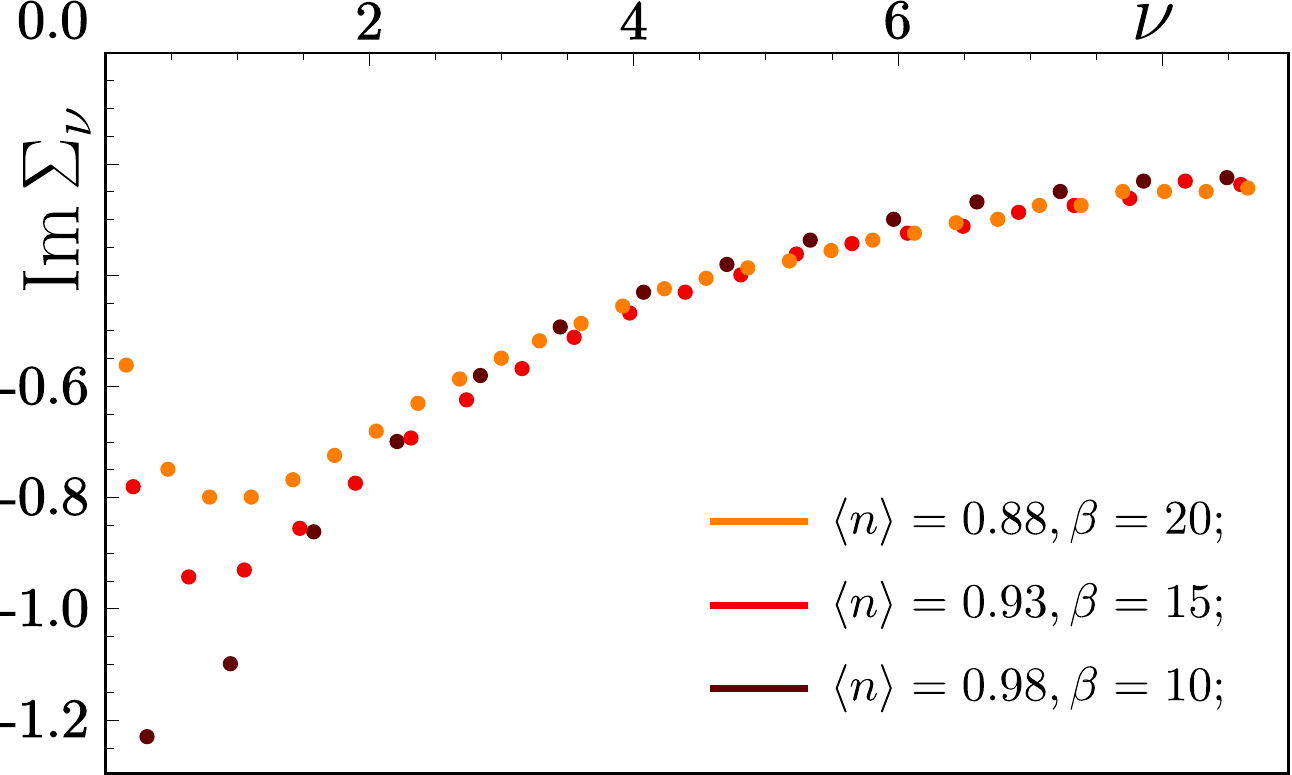}
\end{center}
\vspace{-0.3cm}
\caption{Imaginary part of the electronic self-energy for the extended Hubbard model for cuprates in 
the Matsubara frequency representation obtained in the paramagnetic regime 
close to the PM to AFM phase boundary for different values of hole-doping 
$\av{n}=0.98$, 0.93 and 0.88. The energy and Matsubara frequencies are
given in the eV. 
\label{fig:SM_Sigma}}
\vspace{0.3cm}
\end{figure}

The evolution of the electronic band structure of cuprates with the hole-doping 
is shown in Fig.~\ref{fig:SM_Green}\,a)-d). One can observe that already 
at small hole-doping of 2\,\%, the two-peak structure of the single-particle 
spectral function formed by the lower and upper Hubbard bands transforms to the 
three-peak structure, where the additional quasiparticle resonance appears at 
the Fermi level splitting off from the lower Hubbard band. The further 
increase of the doping to $\av{n}=0.93$ and $\av{n}=0.88$ leads to the 
additional flattening of the quasiparticle band at the vicinity of the nodal 
$\text{M}\Gamma/2$ point and pinning of the Fermi level to the nodal and 
anti-nodal points of the energy spectrum. The redistribution of the 
quasiparticle density upon doping results in sharp peaks and almost 
identical 
behavior of the electronic density at the X and $\text{M}\Gamma/2$ points as 
shown in Fig.~\ref{fig:SM_Green}\,e)-h). The corresponding result for the 
electronic self-energy is presented in the Fig~\ref{fig:SM_Sigma}.

Fig.~\ref{fig:SM_X} shows the magnon spectrum as a function of hole-doping. As one can see, the obtained dispersion of paramagnons is almost unchanged with doping and only reveals progressive broadening with the increase of number of holes. The high intensity at the M point has the maximum at the corresponding energies $E_{\rm max}=67$ meV ($\av{n}=0.98$), $66$ meV ($\av{n}=0.93$) and $61$ meV ($\av{n}=0.88$) and is associated with collective AFM fluctuations.

\end{document}